\documentclass[review, times]{elsarticle}
\usepackage{lineno,hyperref}
\usepackage{amssymb}
\usepackage{amsmath}
\usepackage{tabularx}
\usepackage[english]{babel}
\usepackage{graphicx}
\usepackage{color}
\usepackage{multirow}
\usepackage{subfig} 
\usepackage{bm}
\usepackage{algorithm}
\usepackage{threeparttable}

\newdefinition{rmk}{Remark}
\newdefinition{define}{Definition} 
\newproof{pf}{Proof}

\journal{Elsevier}

\begin{document}

\begin{frontmatter}

\title{A variational-level-set based partitioning method for block-structured meshes}

\author{Shucheng Pan} \ead{shucheng.pan@tum.de}
\author{Xiangyu Hu} \ead{xiangyu.hu@tum.de}
\author{Nikolaus. A. Adams} \ead{nikolaus.adams@tum.de}
\address{Chair of Aerodynamics and Fluid mechanics, Department of Mechanical Engineering, Technical University of Munich, 85748 Garching, Germany}
\begin{abstract}
We propose a numerical method for solving block-structured mesh partitioning problems based on the variational level-set method of (Zhao et al., J Comput Phys 127, 1996) which has been widely used in many partitioning problems such as image segmentation and shape optimization. Here, the variational model and its level-set formulation have been simplified that only one single level-set function is evolved. Thus, the numerical implementation becomes simple, and the computational and memory overhead are significantly alleviated, making this method suitable for solving realistic block-structured mesh partitioning problems where a large number of regions is required. We start to verify this method by a range of two-dimensional and three-dimensional uniform mesh partitioning cases. The results agree with the theoretical solutions very well and converge rapidly. More complex cases, including block-structured adaptive mesh partitioning for single-phase and multi-phase multi-resolution simulations, confirm the accuracy, robustness and good convergence property. The measured CPU time shows that this method is efficient for both two-dimensional and three-dimensional realistic partitioning problems in parallel computing. The proposed method has the potential to be extended to solve other partitioning problems by replacing the energy functional.
\end{abstract}

\end{frontmatter}


\section{Introduction \label{sec:intro}}
Different types of partitioning problems emerge in a large variety of research subject of scientific computing. In image processing, a typical example is the image segmentation technique by which an image is partitioned into multiple pieces to detect the embedded objects. In the spatial multi-scale simulations of e.g. reaction-diffusions processes \cite{yates2015pseudo, flegg2015convergence, spill2015hybrid, harrison2016hybrid}, the computational domain is decomposed into different parts where macroscopic, mesoscopic, and microscopic methods are employed, depending on the species concentration level \cite{robinson2014adaptive}. In network science, the nodes of a complex network are clustered to identify the community structures \cite{newman2004fast}. In addition, large-scale parallel computing, such as direct numerical simulations of Navier-Stokes equations and Motel-Carlo simulations of microscopic processes, requires suitable domain decomposition strategies on massively parallel distributed memory computer architectures. When the finite-difference or finite-volume schemes are used to discretize the governing equations in compuational fluid dynamics, the problem is referred to as mesh partitioning \cite{walshaw2000parallel, korovsec2004solving}.

As partitioning a domain is NP-hard, many heuristic algotithms have been proposed to solve different partitioning problems \cite{hendrickson2000graph}. Particularly for mesh partitioning, traditional methods such as space-filling curve algorithms \cite{sagan2012space} and graph-based algorithms \cite{rantakokko2000partitioning, hendrickson2000graph, diekmann2000shape}, although widely used in the community of scientific computing, either suffer from inaccurate minimizion of data comminication or generates non-connected subdomins \cite{hendrickson2000graph, meyerhenke2009graph}. Another idea, transforming the problem to a partial differential equation (PDE) and solving the PDE by fast algorithms, motives the development of the variational level-set method by Zhao et.al \cite{zhao1996variational} to capture multi-phase motions. The energy functional is formulated with respect to the level-set function and is minimized by solving a level-set advection equation \cite{osher1988fronts}. Sucessful applications of this method include the shape optimization problem where fluid dynamics dependent cost functional \cite{sethian2000structural, osher2001level, wang2003level, allaire2004structural, zhou2008variational, zhou2010level, deng2011topology} is minimized, and the image segmentation problems \cite{chan2001active, vese2002multiphase, li2005level, li2010distance} where the zero level-set contour is evolved to approach the image objects. This method shows better flexibility and robustness than graph partitioning based method. Regarding the mesh partitioning problem, although a variational model and the correspoding level-set formualtion have been proposed in Ref. \cite{zhao1996variational}, it is computational costly as for a large-scale parallel computation the number of level-set advection equations required by this method is large, which limits the application in realistic mesh partitioning. 

In this paper we develop a partitioning method based on the variational concept with Ref. \cite{zhao1996variational} but differs in model formulation and the numerical discretization. We focus on the block-structured mesh partitioning which is simpler than other partitioning problems and has two main objectives, (i) the computational workload at each computing processor is required to identical, namely load balancing condition, (ii) inter-processor data communication is minimized. Unlike the idea of multiple level-set method used previously, our model rather is solved by the regional level-set method \cite{zheng2009simulation, pan2017high, pan2018high}, which means only one single scalar function are evolved, irrespective with the number of regions. Compared to other mesh partitioning methods, the method is more general in the sense that its application is easily extended to other types of partitioning problems. The advantage of this method over other graph-based or particle-based partitioning methods such as the centroidal Voronoi particle method \cite{fu2017physics} is that the communication cost and workload are better measured as the interface location is easily reconstructed by the level-set function, leading to a more accurate method.
The paper is organized as follows. Sec. \ref{sec:num} gives a brief overview of the varitional model of Zhao et. al. \cite{zhao1996variational} and details 
our efficient partitioning method based on this model. We assess the capabilities of the model and it numerical discretization by a ranges of two-dimensional (2D) and three-dimensional (3D) mesh partitioning cases in Sec. \ref{sec:example}. Concluding remarks and outlooks are given in Sec. \ref{sec:con}.

\section{Numerical method} \label{sec:num}
\subsection{Variational model for mesh partitioning \label{sec:model}}
Following the variational model proposed by Zhao et.al. \cite{zhao1996variational}, we consider the solution of the mesh partitioning problem is determined by an evolving interface $C$ in $\mathbb{R}^d$, where $d$ is the dimension number. The interface separates the domain $\Omega$ into $N$ regions, and each region $\Omega_n$ ($n \in [1,N]$) contains $M_n$ computational units (grid points for finite difference schemes or cells for finite volume schemes). The units belonging to the same region are computed by one single CPU processor. We seek the optimal partitioning strategy which balances the workload on every processor and minimizes the total communication volume. This can be considered as a variational problem where energy functional is formulated with respect to workload and communication cost. Similar to Ref. \cite{zhao1996variational}, we formulate the variational problem in the following way. First the energy functional related to the communication cost is defined as
\begin{equation}\label{energy1}
E_s (C) = \frac{1}{2} \sum_{r=1}^{N} \int_{\partial \Omega_r} \rho(\mathbf{x})\, ds, \quad \mathbf{x} \in \partial \Omega_r
\end{equation}
where the density $\rho$ is the communication cost per unit area. We solve the following optimization problem,
\begin{equation}\label{optimization}
\min_{C} E_s (C) \quad \text{subject to} \quad \forall r,\, \int_{\Omega_r} \varrho(\mathbf{x}) \, dv - \bar{m} = 0, \quad \bar{m} = \frac{m}{N}
\end{equation}
where the density $\varrho$ is the computational cost per unit volume and $m = \int_{\Omega} \varrho(\mathbf{x}) \, dv$ is the total computational cost. Then we define the load imbalance induced energy
\begin{equation}\label{energy2}
E_b (C) = \sum_{r=1}^{N} \int_{\Omega_r} \left( m_r -  \bar{m}\right)^2   \, dv = \sum_{r=1}^{N} \left( \int_{\Omega_r} \varrho(\mathbf{x}) \, dv -  \bar{m}\right)^2 
\end{equation}
and minimize the total energy $E$, 
\begin{equation}\label{energytotal}
C^* = \underset{C}{\text{argmin}}\, E(C) = \underset{C}{\text{argmin}} \, \left(  E_s (C) + \mu E_b (C) \right),
\end{equation}
where the parameter $\mu$ is the weight for $E_s$ and $E_b$, similar to that for image segmentation \cite{chan2001active}. In the following, we call $E_s$ and $E_b$ the surface energy and bulk energy, respectively. The combination of communication cost minimization effect and the load balancing consraint is different with that in Ref. \cite{zhao1996variational} where a Lagrangian multiplier is used to combine the two effects.
\subsection{Level-set formulation of the model}\label{sec:pp}
The interface is represented by $C = \left\lbrace \mathbf{x} | \phi(\mathbf{x}) = 0 \right\rbrace $ in the level-set method \cite{osher1988fronts, gibou2017}, where $\phi: \mathbb{R}^d \rightarrow \mathbb{R}$ is the level-set function. Like the Chan-Vese model \cite{chan2001active} for image segmentation, the variational model for mesh partitioning has a level-set formulation. If we use the multiple level-set method in Ref. \cite{zhao1996variational} where $N$ level-set functions are assigned to $N$ regions, the total energy is a functional of all $N$ level-set functions which are solved simultaneously and coupled by Lagrangian multipliers \cite{zhao1996variational}.
However, in a typical large scale parallel computing, the number of CPU processors, $N$, can be hundreds or thousands, which limits the application of method in Ref. \cite{zhao1996variational}. Here we rather employ the regional level-set function \cite{zheng2009simulation, kim2010multi} $\varphi^{\chi} : \mathbb{R}^d \rightarrow \mathbb{R}\times \mathbb{N}$, which is defined as $\varphi^{\chi}(\mathbf{x})=(\varphi(\mathbf{x}), \chi(\mathbf{x}))$, where $\varphi(\mathbf{x}) \geqslant 0$ and $\chi(\mathbf{x})$ is an integer region indicator. Then the energy functional $E(C)$ is rewritten as
\begin{equation}\label{energy}
E(\varphi^{\chi}) = \frac{1}{2} \sum_{r=1}^{N} \int_{\Omega} \rho\, \delta(\phi^r) \vert  \triangledown \phi^r \vert \, dv + \mu \sum_{r=1}^{N} 
\left( \int_{\Omega} H(\phi^r) \varrho \, dv -  \bar{m}\right)^2,
\end{equation}
where $H(\phi)$ is the Heaviside function,  $\delta(\phi)$ is the Dirac delta function and $\bar{m}$ is the averaged workload. The local signed value $\phi^r$ is obtained by a simple mapping $\mathbf{C}_r : \mathbb{R}\times \mathbb{N} \rightarrow \mathbb{R}$ defined as
\begin{equation}\label{eq:map}
\phi^{r}(\mathbf{x})= \mathbf{C}_r \left(\varphi^{\chi}(\mathbf{x})\right) = \begin{cases}
\varphi(\mathbf{x}) \quad {\rm if}~ \chi(\mathbf{x}) = r \cr
-\varphi(\mathbf{x}) \quad \rm{otherwise}	
\end{cases}.
\end{equation}
The G\^{a}teaus derivative of $E(\varphi^{\chi})$ with respect to $\varphi^{\chi}$ is 
\begin{eqnarray}\label{evolve}
\frac{\partial E(\varphi^{\chi})}{\partial \varphi^{\chi}} &=& \sum_{r=1}^{N} \frac{\partial E^r(\varphi^{r})}{\partial \varphi^{r}} = - \frac{1}{2} \sum_{r=1}^{N} \int_{\Omega_r} \delta(\phi^r) \left( \rho  \triangledown \cdot \left( \frac{\triangledown \phi^r}{\vert \triangledown \phi^r \vert} \right) + \frac{\triangledown \rho \cdot \triangledown \phi^r}{\vert \triangledown \phi^r \vert}\right) \, dv \nonumber \\ 
&+& \sum_{r=1}^{N} \int_{\partial \Omega} \frac{\delta(\phi^r)}{\vert \triangledown \phi^r \vert} \frac{\partial \phi^r}{\partial \mathbf{n}}\, ds 
+ 2 \mu \sum_{r=1}^{N} \left(  m_r - \bar{m} \right) \int_{\Omega_r} \varrho \delta(\phi^r) \, dv ,
\end{eqnarray}
where 
\begin{equation}\label{mass}
m_r = \int_{\Omega} \varrho \, H(\phi^r) \, dv
\end{equation}
is the total computational cost of the region $r$ and $m_r - \bar{m}$ is the corresponding load imbalance error. Here we simply call $m_r$ mass and refer  $m_r - \bar{m}$ to mass error.
We obtain the Euler-Lagrangian equation by advecting $\varphi^{\chi}$ in the opposite direction of the G\^{a}teaus derivative and imposing the Neumann boundary conditions,
\begin{equation}\label{evolve}
\frac{\partial \varphi^{\chi}}{\partial t} = \sum_{r=1}^{N} \delta(\phi^r) \left( \rho \triangledown \cdot \left( \frac{\triangledown \phi^r}{\vert \triangledown \phi^r \vert} \right) + \frac{\triangledown \rho \cdot \triangledown \phi^r}{\vert \triangledown \phi^r \vert}\right) - \sum_{r=1}^{N} \mu_r \left(  m_r - \bar{m} \right) \varrho \delta(\phi^r),
\end{equation}
where the weighting factors $\mu_r$ are set as 
\begin{equation}\label{mu}
\mu_r = \frac{c}{\int_{\Omega} \rho\, \delta(\phi^r) \vert  \triangledown \phi^r \vert \, dv}, 
\end{equation}
and $c$ is a constant to ensure the effect of two energy terms in Eq. (\ref{energy}) are in the same order of magnitude, i.e., 
\begin{equation}\label{mu}
\left( \int_{\Omega} \rho\, \delta(\phi^r) \vert  \triangledown \phi^r \vert \, dv\right)  ^2 \simeq  \left( \int_{\Omega} H(\phi^r) \varrho \, dv -  \bar{m}\right)^2.
\end{equation}
After applying the rescale step in Ref. \cite{zhao1996variational}, the Dirac delta function $\delta(\phi^r)$ in Eq. (\ref{evolve}) is replaced by $\vert \triangledown \phi^r \vert$ and Eq. (\ref{evolve}) becomes
\begin{equation}\label{evolve2}
\frac{\partial \varphi^{\chi}}{\partial t} = \sum_{r=1}^{N} \vert \triangledown \phi^r \vert \left( \rho \triangledown \cdot \left( \frac{\triangledown \phi^r}{\vert \triangledown \phi^r \vert} \right) + \frac{\triangledown \rho \cdot \triangledown \phi^r}{\vert \triangledown \phi^r \vert} + \frac{c\, \varrho \left(  \bar{m} -m_r \right) }{\int_{\Omega} \rho\, \delta(\phi^r) \vert  \triangledown \phi^r \vert \, dv} \right)
\end{equation}
which can be easily solved by existing numerical schemes for Hamilton-Jacobi equations \cite{osher1988fronts}.

\subsection{Numerical discretization}\label{sec:pp}
We use the numerical method in Ref. \cite{pan2017high} to efficiently solve the advection equation Eq. (\ref{evolve2}). For problems with large $N$, the evaluation of right hand side terms is costly. Note that before the rescale step most of $\delta(\phi^r)$ in Eq. (\ref{evolve}) will vanish, which simplifies Eq. (\ref{evolve2}). In a 2D Cartesian mesh, we discretize Wq. (\ref{evolve2}) by
\begin{equation}\label{evolve3}
\frac{\partial \varphi^{\chi}_{i,j}}{\partial t} = \sum_{r=1}^{N_s} \vert \triangledown \phi^r \vert_{i,j} \left( \rho_{i,j} \triangledown \cdot \left( \frac{\triangledown \phi^r}{\vert \triangledown \phi^r \vert} \right)_{i,j} + \frac{(\triangledown \rho)_{i,j} \cdot (\triangledown \phi^r)_{i,j}}{\vert \triangledown \phi^r \vert_{i,j}} + \frac{c\, \varrho_{i,j} \left(  \bar{m} -m_r \right) }{\int_{\Omega} \rho\, \delta(\phi^r) \vert  \triangledown \phi^r \vert \, dv} \right), 
\end{equation}
where $(i,j)$ index the grid points and $N_s \ll N$ is the number of regions existing in a local small stencil, e.g. $\{(k,l)|i-1<k<i+1, j-1<l<j+1\}$ in Ref. \cite{pan2017high, pan2018high}. Then we reformulate Eq. (\ref{evolve3}) by
\begin{equation}\label{evolve4}
\frac{\partial \varphi^{\chi}_{i,j}}{\partial t} = \sum_{r=1}^{N_s} u^{r}_{i,j} \vert \triangledown \phi^r \vert_{i,j}
= u_{i,j} \vert \triangledown \phi^{\chi_{i,j}} \vert_{i,j}, 
\end{equation}
where the driven velocity $u^r_{i,j}$ is only computed at near-interface grid points, $\{(i,j) \vert \chi_{i,j} \neq \chi_{i+1,j} \vee \chi_{i,j} \neq \chi_{i-1,j} \vee \chi_{i,j} \neq \chi_{i,j+1} \vee \chi_{i,j} \neq \chi_{i,j-1} \}$. Then the interface velocity is extended to the reset grid points by using the closest point method \cite{ruuth2008simple}.
In this way, the velocity along the interface normal direction remains approximately invariant \cite{ruuth2008simple}. The semi-discrete form of Eq. (\ref{evolve4}) is 
\begin{equation} \label{HJ4}
\frac{\partial \varphi}{\partial t} + H^G (D_x^+ \phi^r, D_x^- \phi^r, D_y^+ \phi^r, D_y^-\phi^r) = 0
\end{equation}
where $H^G(\varphi)$ is the Godunov numerical Hamiltonian \cite{osher1988fronts},
\begin{eqnarray} \label{godunov}
H^G (a,b,c,d) = 
\begin{cases}
\sqrt{\max(\vert a^- \vert^2, \vert b^+ \vert^2) +\max(\vert c^- \vert^2, \vert d^+ \vert^2)} & \quad \text{if $u>0$}\\
\sqrt{\max(\vert a^+ \vert^2, \vert b^- \vert^2) +\max(\vert c^+ \vert^2, \vert d^- \vert^2)} & \quad \text{otherwise},
\end{cases}
\end{eqnarray}
with $f^+ = \max(f,0)$ and $f^- = \min(f,0)$. The derivatives, $D_x^+ \phi^r$, $D_x^- \phi^r$, $D_y^+ \phi^r$ and $D_y^- \phi^r$, are obtained by the 1st-order finite difference scheme. After advection, a remapping $\mathbf{R} : \mathbb{R}^{N_{s}} \rightarrow \mathbb{R} \times \mathbb{N}$,
\begin{equation}
\varphi^{\chi}_{ij} = \mathbf{R}(\phi^{r_1}_{ij},\phi^{r_2}_{ij}, \ldots, \phi^{r_{N_{s}}}_{ij}) = \left( \vert \underset{r}{\max} \phi_{ij}^r \vert, \underset{r}{\text{arg\,max}}\phi_{ij}^r \right)
\end{equation}
is used to reconstruct the regional level-set function, see Ref. \cite{pan2017high, pan2018high} for details.
Moreover, we observe that the Heaviside function in Eq.(\ref{mass}) leads to large oscillations and bad convergence feature, see Fig. \ref{fig:2region_line}. Thus we modify Eq. (\ref{mass}) by 
\begin{equation}\label{eq:mass}
m_r = \int_{\Omega_r} \varrho \, \alpha(\phi^r) \, dv,
\end{equation}
where the volume fraction $\alpha(\phi^r)$ is calculated by geometrical reconstruction such as that in Ref. \cite{lauer2012numerical}.

\subsection{Main features} \label{sec:lts}
Before validate our mesh partitioning method, we first summarize the main features of this method as
\begin{itemize}
\item This method requires less computational and memory cost than that of Ref. \cite{zhao1996variational}, as only one single funtion is evolved. The efficiency of this method can be further increased by the narrow-band technique \cite{peng1999pde} and adaptive mesh refinement techinque \cite{min2007second, mirzadeh2016parallel}. The time-step constraint due to diffusion terms of Eq. (\ref{evolve3}) can be less restrictive by using some operator splitting strategies \cite{goldenberg2001fast, gibou2005fast, estellers2012efficient} and the costly redistancing procedure can be avoided by adding a penalty term \cite{li2005level, li2010distance}. Although additional computational efficiency can be achieved by using those algorithms, the splitting errors and penalty terms may effect the partitioning results. This will be our future work.
\item Usually in the test cases below, we randomly initialize the $\varphi^{\chi}$ field that its energy largely deviates from the equilibrium state. However, in a realistic parallel computing, as the initial $\varphi^{\chi}$ is chosen to be the equilibrium state of the previous partitioning result and the flow field changes continuously in time, the iteration number becomes smaller.
\item Like the original level-set method \cite{osher1988fronts}, the method is well suitable to be parallized, and the computational cost and communicated data are neglectable compared to that of the fluid dynamic simulations, as the level-set data is defined on a coarser resolution, e.g. the block level.
\item Apart from the mesh partitioning for parallel computing, this method can be applied to other partitioning problems by suitably formulate the energy functional in Eq.(\ref{energy}). For example, to solve the image segmentation problem, one can replace the bulk energy by an energy based on image gray scale, like the Chan-Vese model \cite{chan2001active}. For the spatial multi-scale simulations of reaction-diffusions processes \cite{yates2015pseudo, flegg2015convergence, spill2015hybrid, harrison2016hybrid}, one can define the energy term according on the concentration of species and dynamically partition the domain \cite{robinson2014adaptive}.
\end{itemize}
\section{Numerical examples}\label{sec:example}

In this section, we apply our numerical method to solve a number of 2D and 3D mesh partitioning cases. The 1st-order upwind scheme and the explicit Euler scheme are used for spatial discretization and time marching, respectively. If not mentioned otherwise, the CFL number is $0.6$ and the parameter $c$ is $1.0$.

\subsection{Simple test cases}
\subsubsection{2D cases}
\begin{figure}[h]
\begin{center}
\includegraphics[width=1.0\textwidth]{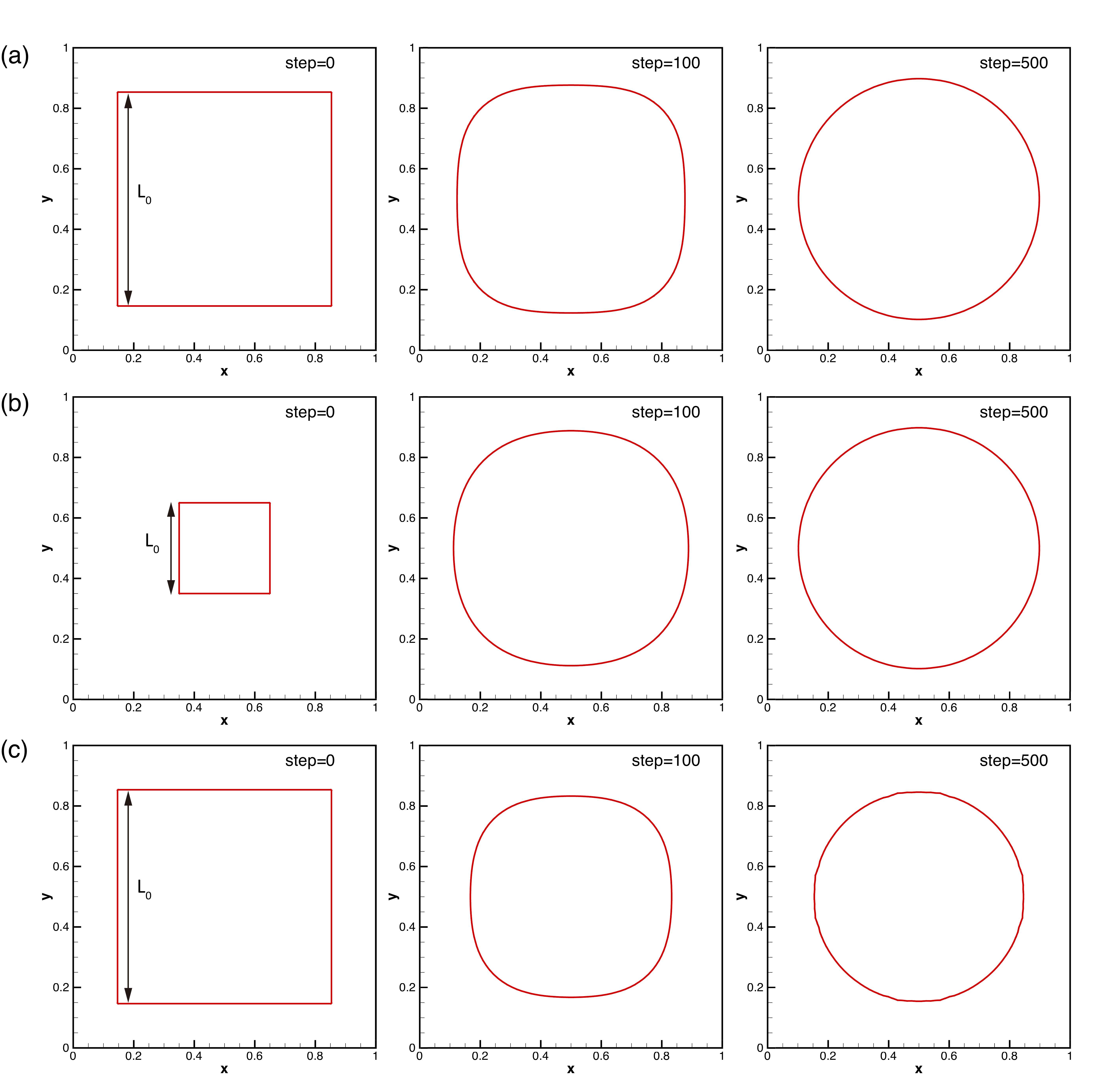}
\caption{Interface evolution of three 2D 2-region mesh partitioning problems: (a) $L_0=1/\sqrt{2}$ and $\varrho=1$ (case 1), (b) $L_0=0.3$ and $\varrho=1$ (case 2), and (c) $L_0=1/\sqrt{2}$, $\varrho=1$ (inside), and $\varrho=0.5$ (outside) (case 3). $\rho=1$ for all cases. The resolution for level-set advection in each case is set as $64 \times 64$ which corresponds to $6400 \times 6400$ total mesh grid points if every block contains $100\times 100$ grid points. See Movie 1 in the supplementary material for details.}
\label{fig:2regioncontour}
\end{center}
\end{figure}

\begin{figure}[h]
\begin{center}
\includegraphics[width=0.8\textwidth]{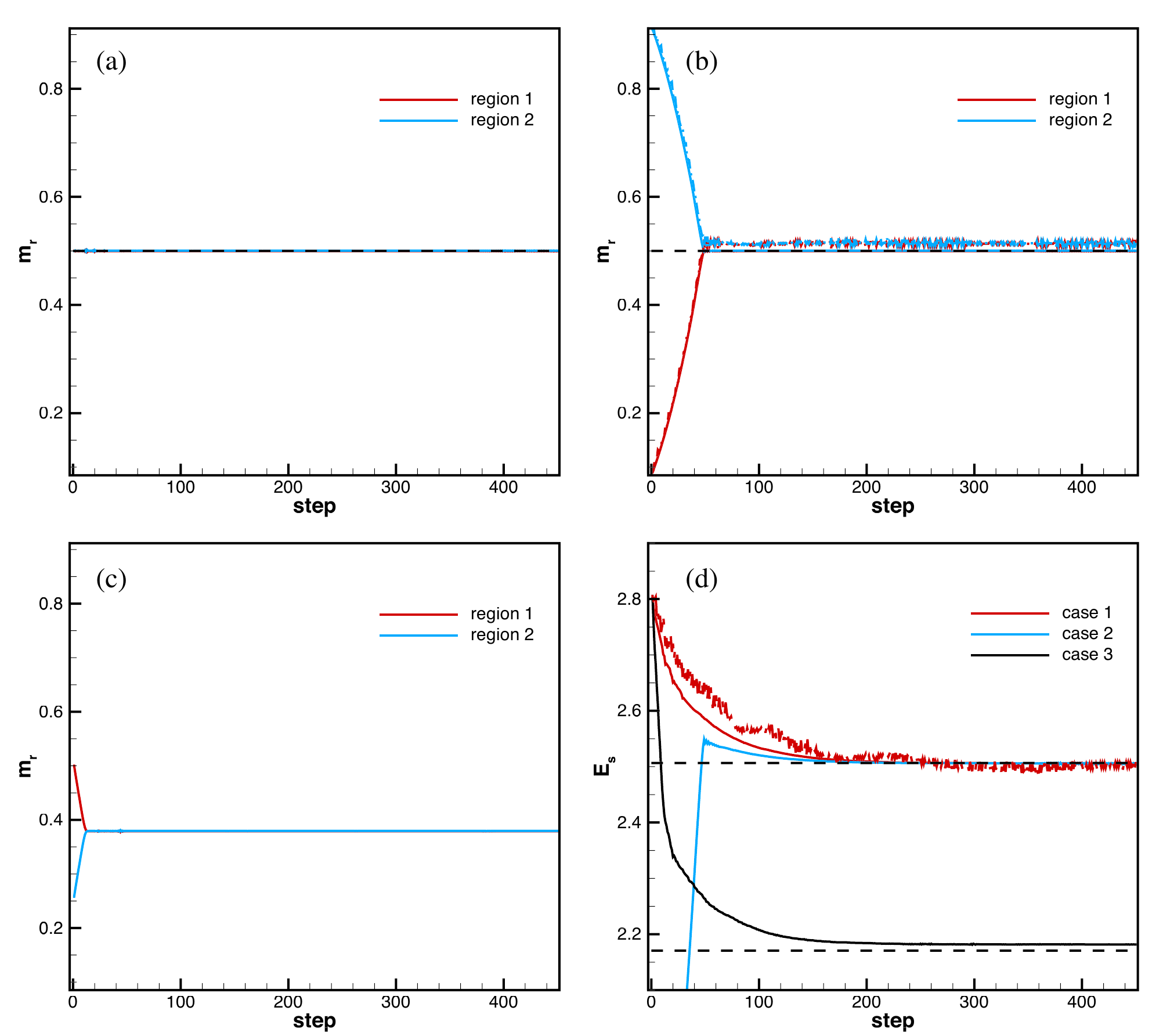}
\caption{Time history of three 2D 2-region mesh partitioning results in Fig. \ref{fig:2regioncontour}. The mass of each region $m_r$ of cases 1, 2 and 3 are show in (a), (b) and (c), respectively. (d) shows the surface energy $E_s$ of each case. The dashed lines indicate the theoretical steady states. The dash-dotted lines in (b) and (d) are the results by using Heaviside function in Ref. \cite{zhao1996variational}.}
\label{fig:2region_line}
\end{center}
\end{figure}

We start with simple mesh partitioning problems where theoretical solution exists. First we consider a two-region case where the density functions are constant, $\varrho=1$ and $\rho=1$. This turns out to be the classic minimal surface problem under the volume-preservation constraint \cite{chopp1993computing, traasdahl2011high}. Case 1, as shown in Fig. \ref{fig:2regioncontour}(a), initially has a load-balancing state (two equal size regions) but the communication volume (total interface length) is larger than the theoretical value. The results of our method show that the square interface continuously becomes circular while the area of two regions keeps identical, as shown in Figs. \ref{fig:2region_line}(a) and (d). The exact mass $m_r$ and surface energy $E_s$ are obtained within about $200$ iterations. Then we consider a case with unequal region area, see case 2 of Fig. \ref{fig:2regioncontour}(b). The square expands and transforms to a circle finally. Fig. \ref{fig:2region_line}(b) shows that the region area imbalance monotonically converges to zero and Fig. \ref{fig:2region_line}(d) shows that the interface length first increase and decreases to the exact solution afterwards. We also consider the non-uniform density distribution for case 3 in Fig. \ref{fig:2regioncontour}(c), $\varrho=1.0$ inside the square and $\varrho=0.5$ outside, i.e., the area of each region is the same but the mass is not. The interface evolution and time history of energy terms are shown in Fig. \ref{fig:2regioncontour}(c) and Fig. \ref{fig:2region_line}(c), respectively. The results fit the theoretical value very well and converges rapidly, as expected. Note that if we use the Heaviside function suggested in Ref. \cite{zhao1996variational}, rather Eq. (\ref{eq:mass}), to compute the bulk energy terms, the results exhibit large oscillation (Figs. \ref{fig:2region_line}(b) and (d)) and may converge to wrong solution (Fig. \ref{fig:2region_line}(b)).

\begin{figure}[h]
\begin{center}
\includegraphics[width=1.0\textwidth]{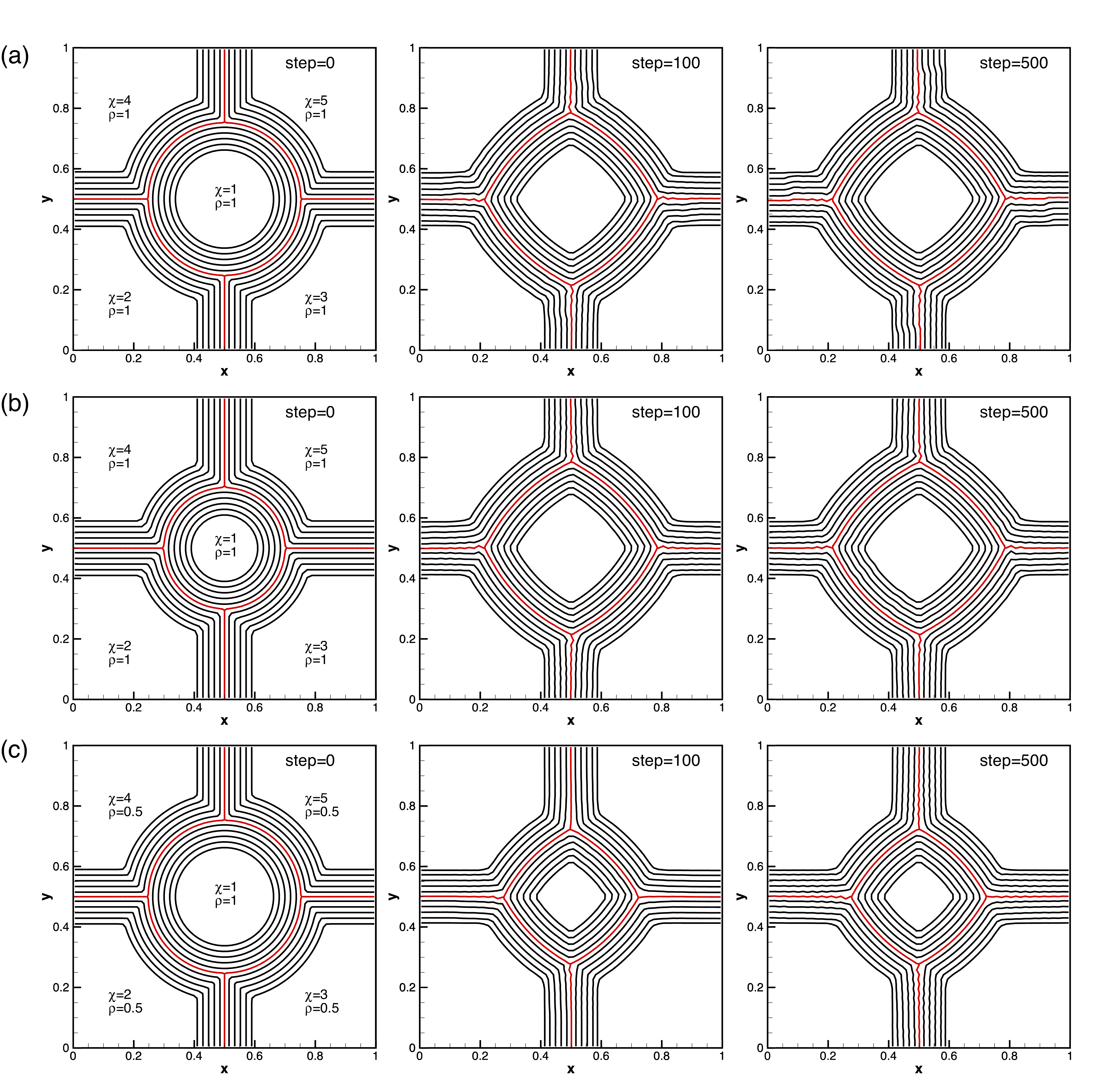}
\caption{Interface evolution of three 2D 5-region mesh partitioning problems: (a) $R_0=1/\sqrt{5 \pi}$ and $\varrho=1$ (case 3), (b) $R_0=0.2$ and $\varrho=1$ (case 4), and (c) $R_0=1/\sqrt{5 \pi}$, $\varrho=1$ ($\chi=1$), and $\varrho=0.5$ ($\chi=2-5$) (case 5), where $R_0$ is the radius of the circular region ($\chi=1$). $\rho=1$ for all cases. The resolution for level-set advection in each case is set as $64 \times 64$. The level-set contour, represented by black lines, ranges from $\varphi=0.15$ to $\varphi=0.9$. The interface is colored by red. See Movie 2 in the supplementary material for details.}
\label{fig:5regioncontour}
\end{center}
\end{figure}

\begin{figure}[h]
\begin{center}
\includegraphics[width=0.9\textwidth]{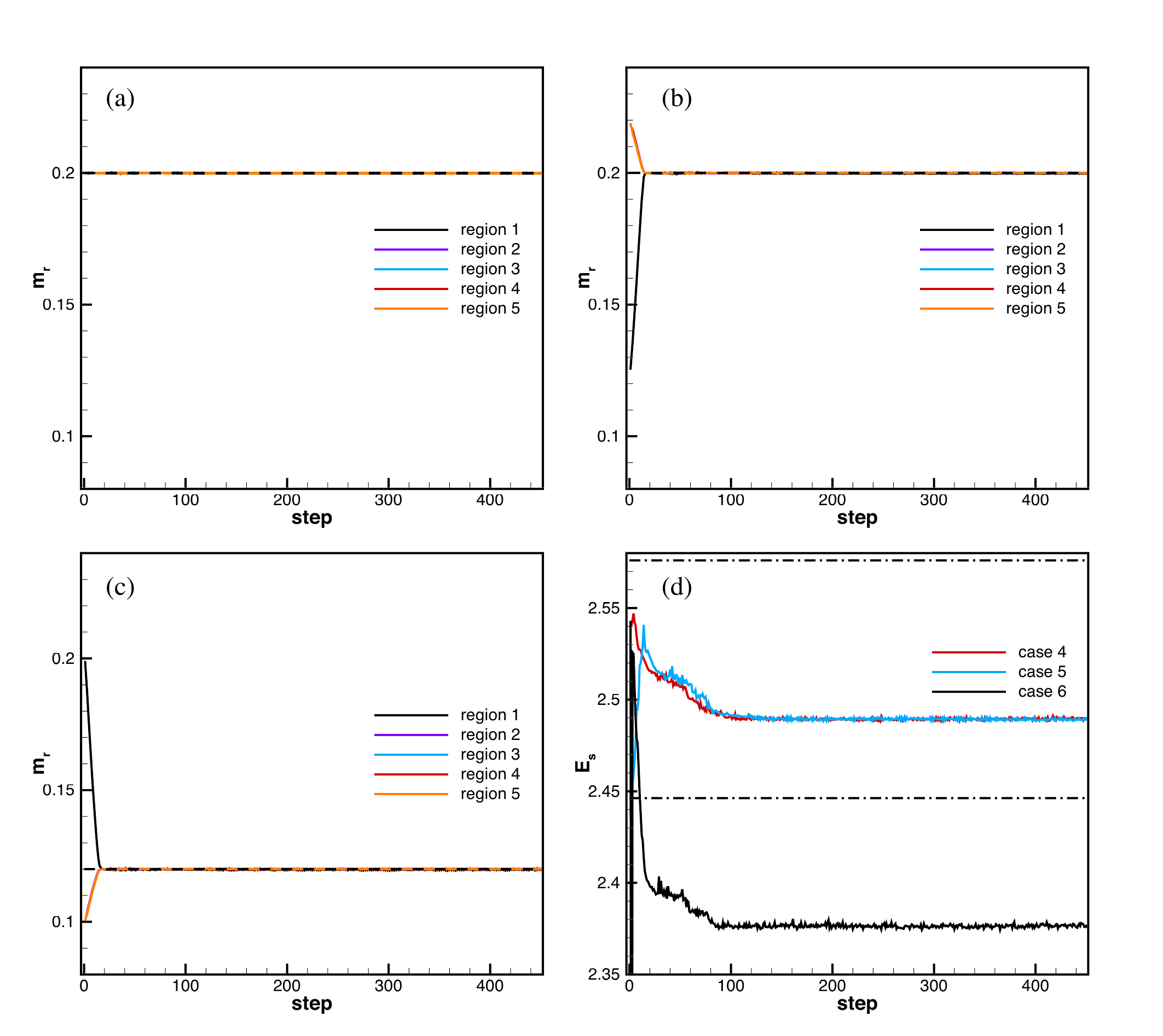}
\caption{Time history of three 2D 5-region mesh partitioning results in Fig. \ref{fig:5regioncontour}. The mass of each region $m_r$ of cases 4, 5 and 6 are show in (a), (b) and (c), respectively. (d) shows the surface energy $E_s$ of each case. The dashed lines indicate the theoretical steady states and dash-dot lines provide the reference communication cost.}
\label{fig:5region_line}
\end{center}
\end{figure}

Then we increase the complexity of partitioning problems by adding triple points in the system. As shown in Fig. \ref{fig:5regioncontour}, the computational domain is decomposed by $5$ regions and the initial conditions of three cases, case 3-5, are shown in the first column. Both uniform (Figs. \ref{fig:5regioncontour}(a) and (b)) and non-uniform (Fig. \ref{fig:5regioncontour}(c)) density distribution are considered. For all three cases, Fig. \ref{fig:5regioncontour} shows that the circular region ($\chi=1$) transforms a curved polygon during iterations and the angles near the triple points converges to approximately $120^{\circ}$. We observe that the interface near the $4$ triple points is not smooth which can be improved by the redistancing-like mapping operator in Ref. \cite{pan2017high}. As this operation is more costly than Eq. (\ref{eq:map}) and the oscillating of triple points location is small, we tolerate this for mesh partitioning.  
The workload of each region $m_r$ converges to the exact value rapidly, as shown in Figs. \ref{fig:5region_line}(a)-(c). The surface energy $E_s$ evolution in Fig. \ref{fig:5region_line}(d) converges to steady value with small fluctuations due to the triple point motion. This steady value of $E_s$ is smaller than a reference solution which corresponding to a 5-region system with identical $m_r$ and a circular inner region ($\chi=1$).

\subsubsection{3D cases}
\begin{figure}[h]
\begin{center}
\includegraphics[width=0.65\textwidth]{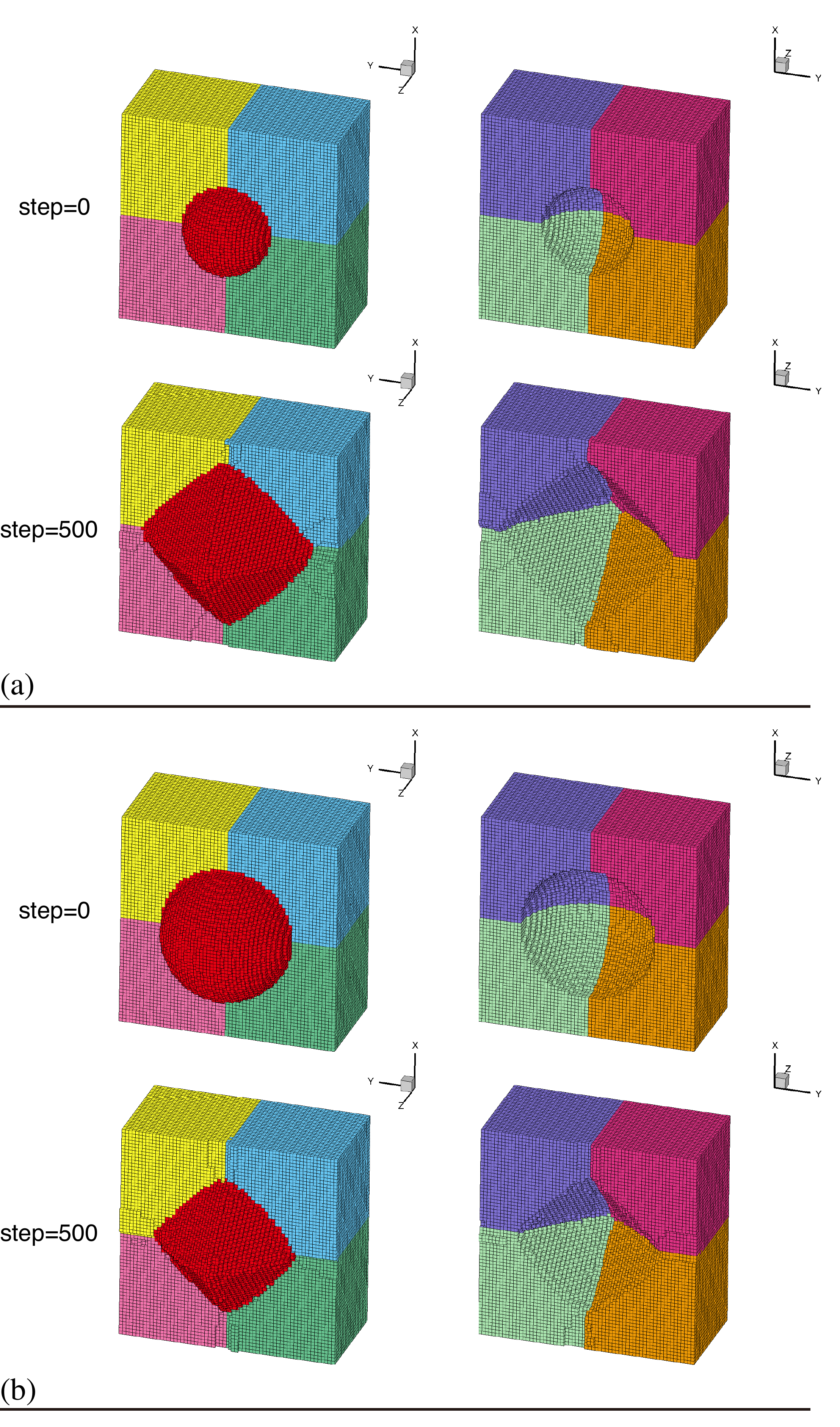}
\caption{Mesh partitioning results of two 3D 9-region mesh partitioning problems: (a) $R_0=0.2$ and $\varrho=1$ (case 7), (b) $R_0=0.3$, $\varrho=1$ for $\chi=1$, and $\varrho=0.5$ for $\chi=2-9$ (case 8), where $R_0$ is the radius of the spherical region ($\chi=1$). $\rho=1$ for all cases. The resolution for level-set advection in each case is set as $64 \times 64 \times 64$.}
\label{fig:9region3dpart}
\end{center}
\end{figure}

\begin{figure}[h]
\begin{center}
\includegraphics[width=0.9\textwidth]{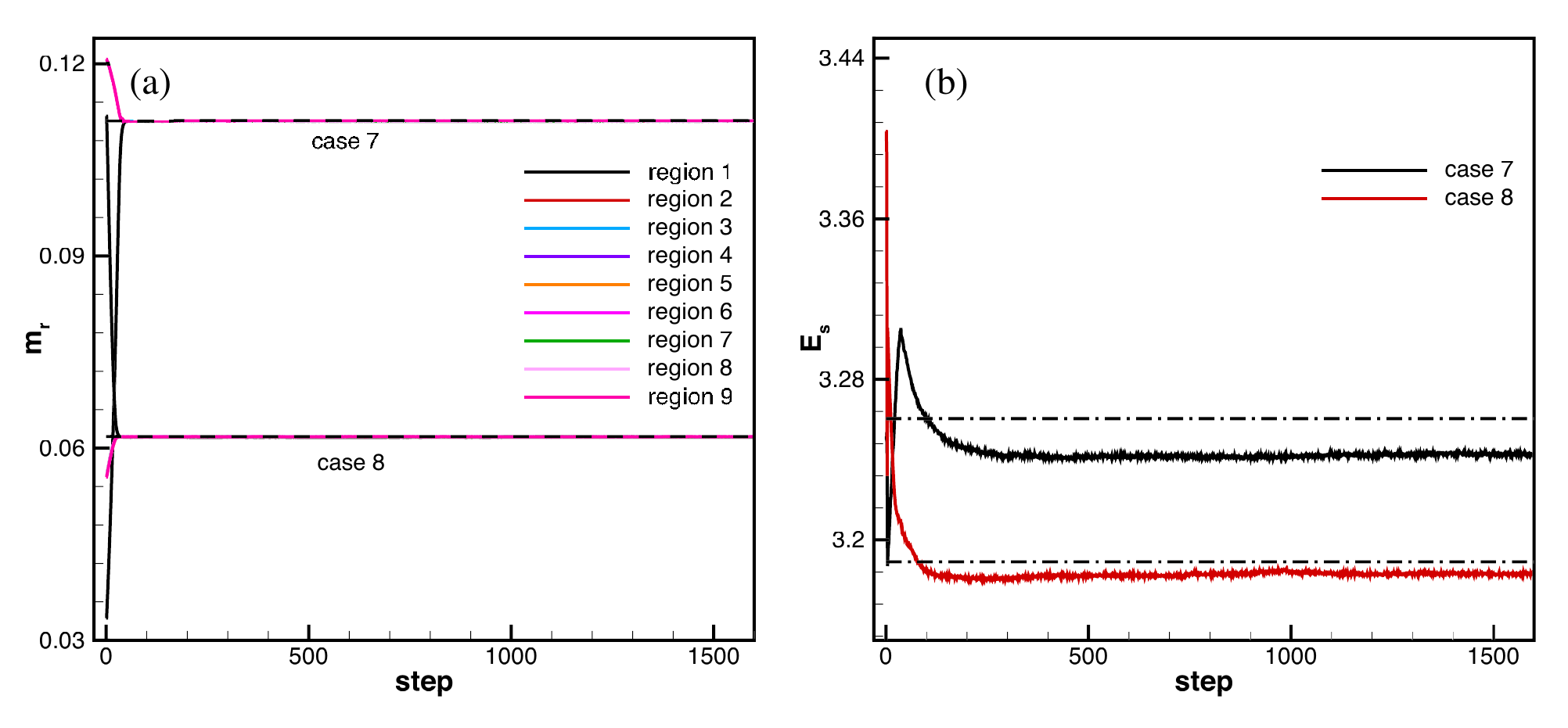}
\caption{Time history of mass (a) and surface energy (b) for two 3D 9-region mesh partitioning results in Fig. \ref{fig:9region3dpart}. The dashed lines indicate the theoretical steady states and dash-dot lines provide the reference communication cost.}
\label{fig:9region3d}
\end{center}
\end{figure}

Here we extend the 5-region 2D case to 9-region 3D case. The domain $[0,1]\times[0,1]\times[0,1]$ is decomposed into $9$ regions: a sphere ($\chi=1$) centered at $(0.5, 0.5, 0.5)$ with a radius of $R_0$ and the exterior of the sphere which is equally split by $8$ regions ($\chi=2-9$). We consider two configurations, i.e., case 7 with $\varrho=1$ and $R_0=0.2$, and case 8 with $\varrho=1$ and $R_0=0.2$. This indicates that the initial state does not satisfy the load balancing condition. In both cases, the interface is evolved to reach the minimal surface energy $E_s$ under the constraint of load balancing. Similar to 2D cases, the spherical region deforms to curved polyhedra, see Fig. \ref{fig:9region3dpart}. The initial load imbalance is decreased to zero and the mass of each regions agrees with the exact value, as shown in Fig. \ref{fig:9region3d}(a). The surface energy is minimized quickly below the reference value which corresponding to a topology composed of $9$ identical $m_r$ and a spherical inner region ($\chi=1$).

\subsection{Adaptive mesh partitioning}
After validating our method by simple 2D and 3D uniform mesh partitioning problems, we present adaptive mesh partitioning cases in this section. We consider both single-phase and multi-phase compressible flows simulations with adaptive mesh refinement technique. If the numerical simulations for those flows utilize $\ell_{\max}$ different resolution levels, we define the density field according to the local mesh level $\ell(\mathbf{x})$, $\varrho(\mathbf{x})=2^{1-\ell(\mathbf{x})/\ell_{\max}}$, and set $\rho=\varrho$.
\subsubsection{Single-phase simulations}
\begin{figure}[h]
\begin{center}
\includegraphics[width=1.0\textwidth]{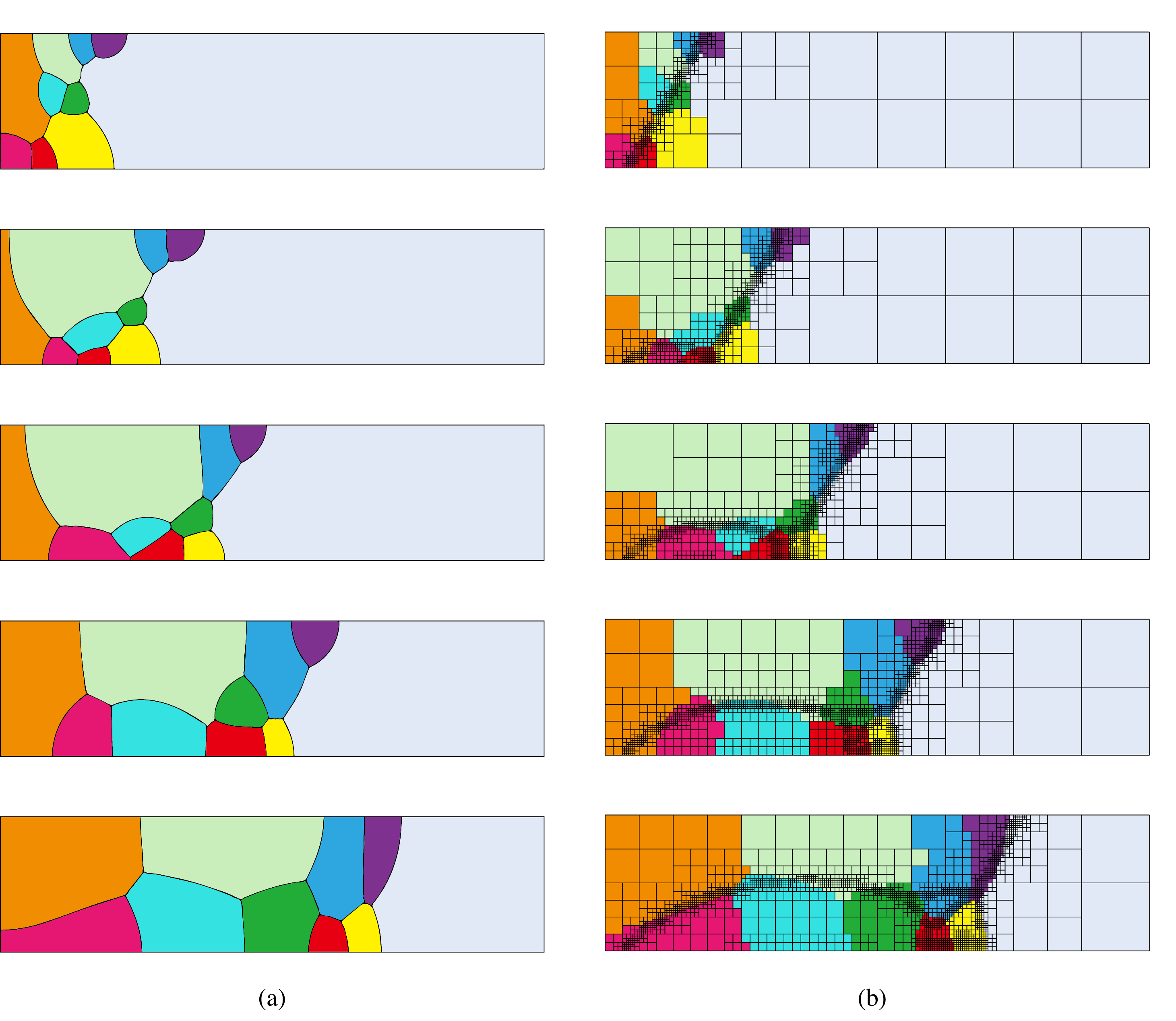}
\caption{The interface at steady state (a) and the corresponding 10-region partitioning results (b) for 2D multi-resolution simulations of double Mach reflection at $5$ physical time instants $t=0$, $t=0.05$, $t=0.1$, $t=0.15$ and $t=0.2$ (from the top row to the bottom row). The resolution for level-set advection in each case is set as $256 \times 64$ and the effective resolution for the flow field is $4096 \times 1024$.}
\label{fig:doublemachpart}
\end{center}
\end{figure}

The double Mach reflection with a Mach $10$ shock wave is chosen in this section to demonstrate our mesh partitioning method for adaptive single-phase parallel simulations, see Ref. \cite{colella1984piecewise} for problem description and initial condition. We use the multi-resolution solver \cite{han2011wavelet} to generate flow fields and block distributions which will be used to calculate the background $\varrho$ field. The resolution level is set as $\ell_{\max}=6$ and each block contains $16\times 16$ cells, leading to an effective resolution of $1024\times 4096$. The blocks with finest resolution are mainly located near the shock wave loactions by the multi-resolution representation of Ref. \cite{harten1995multiresolution}, ash shown in Fig. \ref{fig:doublemachpart}(b). The computational domain, $[0,4]\times [0,1]$, is decomposed by $N=10$ randomly generated voronoi cells initially, with each containing unequal computational workload. We test our method at $5$ physical time, $t=0$, $0.05$, $0.1$, $0.15$ and $0.2$. As shown in Fig. \ref{fig:doublemachpart}, after a number of iterations, the straight voronoi edges become curved during the multi-region system motion, leading to a foam-like topology, which is similar to the minimal surface problem \cite{chopp1993computing, traasdahl2011high} which seeks the least possible surface area of a given volume. Thus we consider our method belongs to physical-motived method which mimics the dry-foam dynamics under volume preservation constraints. 

\begin{figure}[h]
\begin{center}
\includegraphics[width=0.8\textwidth]{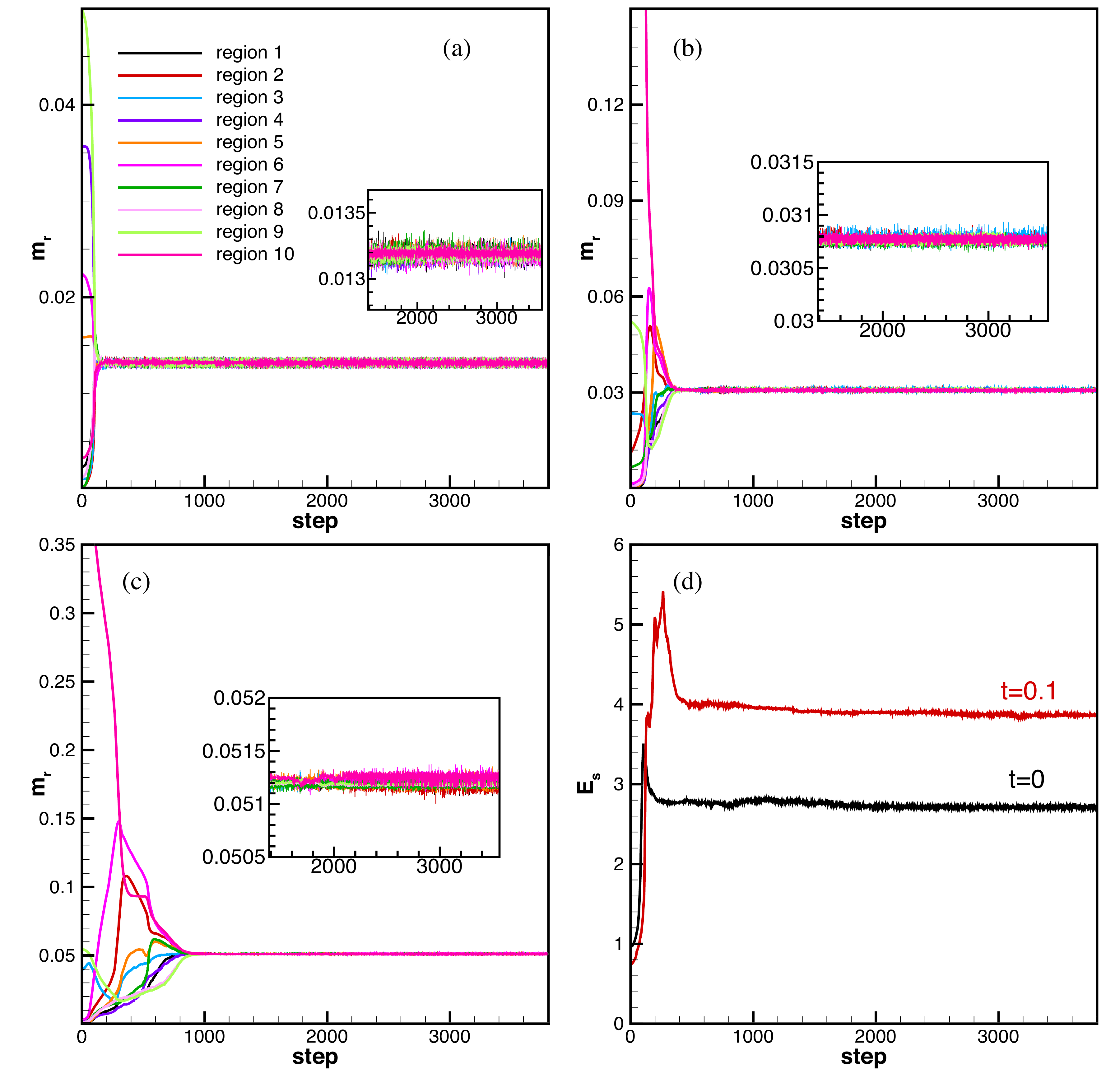}
\caption{Time history of mass (a-c) and surface energy (b) of the 10-region partitioning for 2D multi-resolution simulations of double Mach reflection in Fig. \ref{fig:doublemachpart}.
}
\label{fig:doublemach}
\end{center}
\end{figure}

Here the resolution of level-set advection is identical to the resolution of blocks, $256 \times 64$. When the resolution is not consistent with that of blocks, one need interpolate the density functions, $\varrho$ and $\rho$, which is defined on blocks. We plot the time evolution of region workload $m_r$ for $t=0$, $t=0.1$ and $t=0.2$ in Figs. \ref{fig:doublemach}(a), (b) and (c), respectively. It takes less than 1000 steps to converges to steady solution for each region, with small oscillations shown in the inserts of Fig. \ref{fig:doublemach}. The relative mass errors, $\vert m_r-\bar{m}\vert/\bar{m}$, for each region is less than 1\%, which demonstrates the high accuracy of our method for adaptive mesh partitioning problems. The surface energy $E_s$, representing communication cost, is shown in Fig. \ref{fig:doublemach}(d) and is minimized rapidly.

\subsubsection{Multi-phase simulations}

\begin{figure}[h]
\begin{center}
\includegraphics[width=0.55\textwidth]{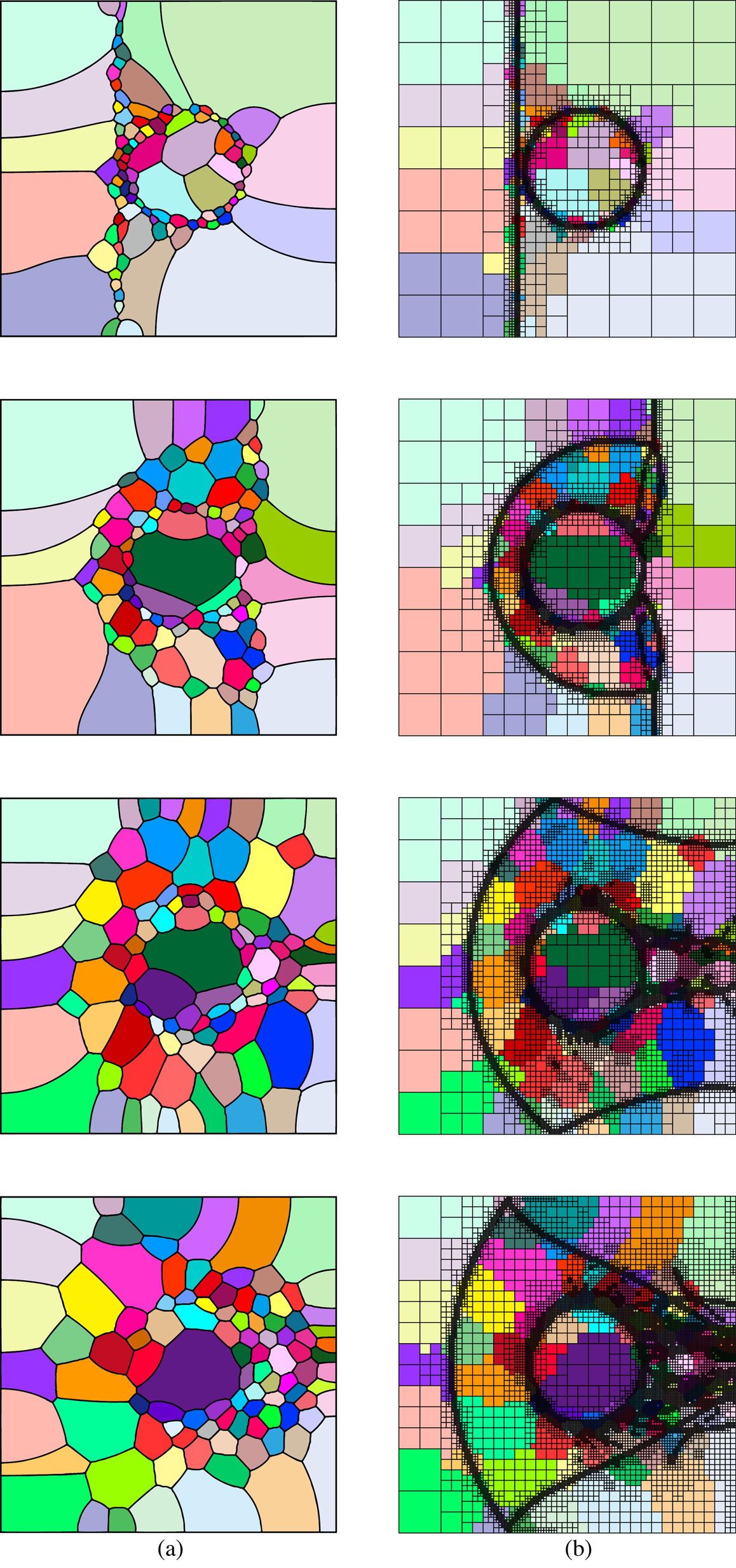}
\caption{The steady interface topologies (a) and the corresponding 100-region partitioning results (b) for 2D multi-resolution simulations of shock-water-column interaction at $4$ physical time instants $t=0$, $t=0.1264$, $t=0.2529$, and $t=0.3794$ (from the top row to the bottom row). The resolution for level-set advection in each case is set as $256 \times 256$ and the effective resolution for the flow field is $4096 \times 4096$. See Movie 3 for details.
}
\label{fig:shockwaterpart}
\end{center}
\end{figure}

\begin{figure}[h]
\begin{center}
\includegraphics[width=0.8\textwidth]{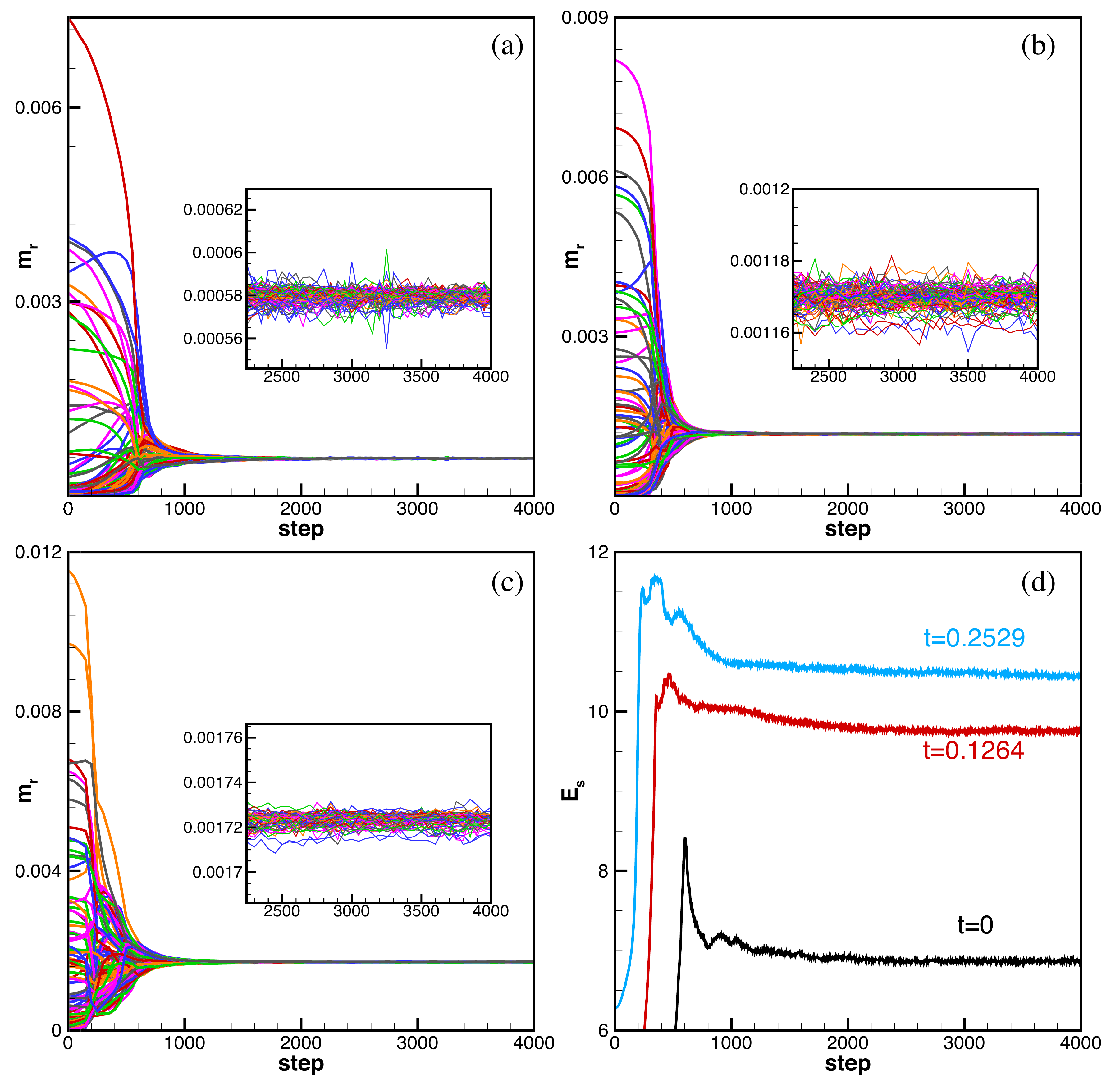}
\caption{
Time history of mass (a-c) and surface energy (b) of the 100-region partitioning for 2D multi-resolution simulations of shock-water-column interaction in Fig. \ref{fig:shockwaterpart}.}
\label{fig:shockwater}
\end{center}
\end{figure}

We consider the shock-water-column interaction problem with a Mach $3$ shock wave \cite{chang2013direct} to test our method for mesh partitioning of multiphase flow simualtions. We use the sharp-interface multi-phase solver \cite{han2014adaptive} which adaptively refines the mesh according to the interface location and shock wave dynamics, as shown in Fig. \ref{fig:shockwaterpart}(b). As the sharp-interface method requires more computational operations (including interface fluxes calculation, flow state extending, flux mixing, level-set advection and redistancing, etc, see Ref. \cite{han2014adaptive} for details) in a multi-phase block than that in a single-phase block, we obtain a density value larger than $1.0$ in those multi-phase blocks, as shown in the last subfigure of Fig. \ref{fig:shockwater}(a). 
The resolution level is set as $\ell_{\max}=8$ and each block contains $16\times 16$ cells, leading to an effective resolution of $4096\times 4096$. The computational domain is a unit square and is decomposed by $N=100$ randomly generated voronoi cells initially. We show our mesh partitioning result at physical times, $t=0$, $0.1264$, $0.2529$, and $0.3794$ in Fig. \ref{fig:shockwaterpart}. The regions concentrate near the shock patterns where the density $\varrho$ is large. The initial large load imbalance, as shown in Figs. \ref{fig:shockwater}(a)-(c), decrease to the steady value very fast. 
The mass error, $\vert m_r - \bar{m}\vert$, as shown in the insert of Fig. \ref{fig:shockwater} is small, which demonstrates the accuracy of our method. The surface energy $E_s$ plotted in Fig. \ref{fig:shockwater}(d) demonstrates the communication cost is minimized under the load balancing constraint.

\begin{figure}[h]
\begin{center}
\includegraphics[width=1.0\textwidth]{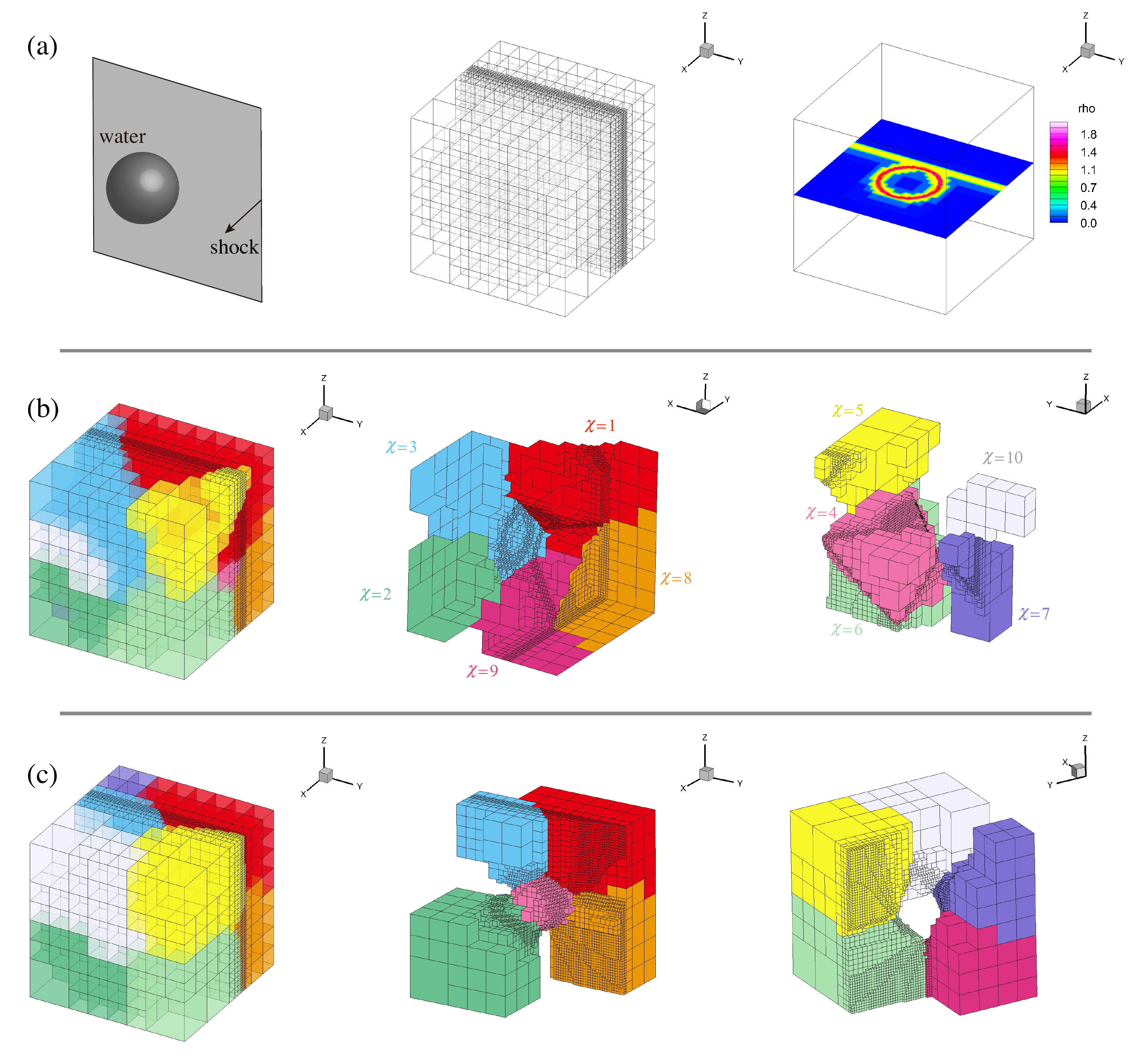}
\caption{(a) The computational configuration of the 3D shock-water-drop interaction: a shock wave of Mach 3 hits a spherical water drop. The block distribution and density field are shown in the middle and right column. The initial and converged mesh partitioning results are shown in (b) and (c), respectively.}
\label{fig:shockwater3dpart}
\end{center}
\end{figure}

\begin{figure}[h]
\begin{center}
\includegraphics[width=0.8\textwidth]{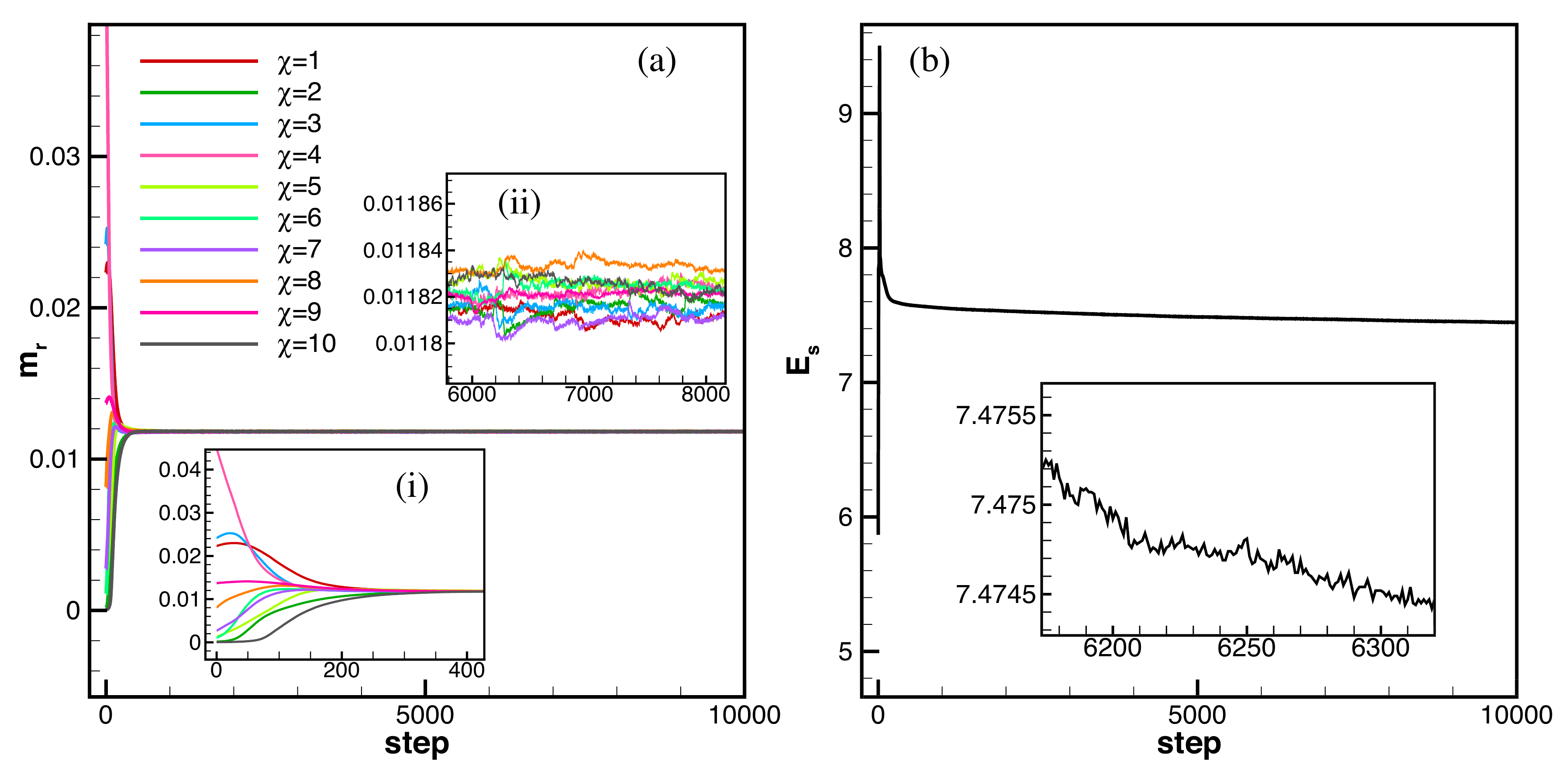}
\caption{
Time history of mass (a-c) and surface energy (b) of the 10-region partitioning for 3D multi-resolution simulations of shock-water-drop interaction in Fig. \ref{fig:shockwater3dpart}.}
\label{fig:shockwater3d}
\end{center}
\end{figure}

Then we consider the 3D shock-water-drop interaction problem with the same initial condition of the above 2D case, see Fig. \ref{fig:shockwater3d}(a) for the problem configuration. We set $\ell_{\max}=5$ and each block contains $16\times 16 \times 16$ cells, corresponding an effective resolution of $1024\times 1024 \times 1024$ for fluid dynamics simulations. The computational domain is randomly partitioned into $N=10$ regions, as shown in Fig. \ref{fig:shockwater3dpart}(b). Clearly, this partition exhibits large load imbalance. For example, the region $\chi=4$ contains a large number of blocks while the region $\chi=10$ merely has $9$ blocks, as qualified in the insert (i) of Fig. \ref{fig:shockwater3d}(a). After achieving the steady solution, we observe the optimal partition in Fig. \ref{fig:shockwater3dpart}(c). Its realtive mass error  $\vert m_r - \bar{m}\vert/\bar{m}$ is less than $0.2\%$, as shown in the insert (ii) of Fig. \ref{fig:shockwater3d}(a). The corresponding surface energy $E_s$ in Fig. \ref{fig:shockwater3d}(b) is minimized within $1000$ steps.

The efficiency of our method is demonstrated by measuring the CPU time for the above mesh partitioning cases. As shown in Table \ref{table_CPU}, our method exhibit high efficiency for the double Mach reflection mesh partitioning problems and only takes $1$ or $2$ seconds on a 8-core (Intel Xeon E5620) desktop. It is about $300$ times faster than another particle-based physical-motived mesh partitioning method \cite{fu2017novel} due to three reasons, (i) our method only uses 1st-order finite difference scheme for advection equation while the interpolation stencil (kernel function) of smoothed particle hydrodynamics (SPH) method of Ref. \cite{fu2017novel} requires more data, especially in 3D, which is costly; (ii) the updated physical state includes particle coordinates, pressure, density and velocity vector while our method only updates one single unsigned level-set function; (iii) traditional schemes \cite{osher1988fronts} for level-set advection exhibits fast convergence speed which is considered as a main numerical issue of SPH method. We note that the 3D mesh partitioning shown in Fig. \ref{fig:shockwater3dpart}, takes about $45$ seconds at one single computing node (20-core Intel Xeon E5-2660 v2), which can be reduced by using more nodes as the nuemrial method for solving level-set advection used here \cite{osher1988fronts} is well suitable for distributed memory parallelization.

\begin{table}
\hspace{-0cm}
\centering
\resizebox{\textwidth}{!}{
\begin{threeparttable}
{\fontsize{10pt}{12pt}\selectfont
\caption{\label{table_CPU} CPU time (in seconds) of mesh partitioning problems for double Mach reflection ($256 \times 64$ grid points) and shock-water interaction ($256 \times 256$ grid points for 2D and $256 \times 256 \times 256$ grid points for 3D) multi-resolution simulations. The 2D simulations and the 3D simulations are performed on a 8-core (Intel Xeon E5620) desktop and a 20-core (Intel Xeon E5-2660 v2) node in a linux cluster of Leibniz Supercomputing Centre, respectively. The wall clock time is measured when both the mass error and surface energy have been minimized. The corresponding iteration number is $1600$ for 2D and $1000$ for 3D cases to ensure the steady solution is achieved.}
\vspace{0.5cm}
\begin{tabular}{ccccccc}
\hline
 & & \small Advection & \small Redistancing & \small Energy terms evaluation & \small Others \tnote{*} & \small Total \\
\hline \multirow{3}*{\small Double Mach reflection} & \small Fig. \ref{fig:doublemach}(a) & \small 13.9\%	& \small 30.4\% & \small 21.6\%		&\small 34.0\% 	&\small 1.77(s)\\
										& \small Fig. \ref{fig:doublemach}(b) & \small 15.0\%	& \small 31.6\% &\small 22.9\%		&\small 30.7\% 	&\small 1.80(s) \\
										& \small Fig. \ref{fig:doublemach}(c) & \small 13.7\% & \small 31.4\% &\small 23.3\%		&\small 31.6\% 	&\small 2.20(s) \\
 \multirow{2}*{\small Shock-Water interaction} 	& \small Fig. \ref{fig:shockwater}(a)  & \small 15.1\% & \small 32.1\%	& \small 18.4\%		& \small 34.4\%	&\small 19.8(s)		\\
 												&\small Fig. \ref{fig:shockwater}(b) &\small 15.7\%	&\small 31.9\% &\small 18.0\%		&\small 34.4\% 	&\small 17.7(s)\\
 												&\small Fig. \ref{fig:shockwater}(b) &\small 15.8\%	&\small 33.0\% &\small 18.4\%		&\small 32.9\% 	&\small 17.2(s)\\
 												&\small Fig. \ref{fig:shockwater3d} &\small 16.8\%	&\small 33.4\% &\small 19.2\%		&\small 30.6\% 	&\small 44.6(s)\\
\hline\\
\end{tabular}
    \begin{tablenotes}
    \small \item[*]{\small This includes computational cost of calculating volume fraction, updating tags, time history of energy terms output and computing time-step, etc.}
    \end{tablenotes}
}
\end{threeparttable}
}
\end{table}

\section{Conclusion and outlook} \label{sec:con}
In this paper we have proposed a variational level-set method for solving mesh partitioning problems encountered in large-scale parallel simulations. We simplify the original variational level-set method by evolving an unsigned level-set function and a region indicator function. In this way, solving the interface evolving equation corresponding to the variational model becomes computational and memory efficient, indicating that the present method is suitable for realistic mesh partitioning problems.
A number of simple 2D and 3D uniform mesh-partitioning cases show that our method can obtain the theoretical results, which demonstrates the accuracy of our method. The load imbalance errors converge to zero and the total communication cost is minimized rapidly. More complex adaptive mesh partitioning problems, including the 2D double Mach reflection, 2D shock-water-column interaction and 3D shock-water-drop interaction simulations, are tested. The results and the CPU time measurement demonstrate that our method is robust, accurate, and efficient.
The coupling with the multi-resolution single-phase or multi-phase simulations will be future work. Its performance on unstructured mesh partitioning and heterogeneous computing platforms needs to be studied. This method has the potential to be applied to other similar problems such as image segmentation  and multi-scale coupling method  by replacing the energy functional.
\section*{Acknowledgment}

This work is supported by China Scholarship Council (No. 201306290030), National Natural Science Foundation of China (No. 11628206), and Deutsche Forschungsgemeinschaft (HU 1527/6-1).
The project has received funding from the European Research Council (ERC) under the European Union's Horizon 2020 research and innovation program (No. 667483). The authors acknowledge Leibniz Supercomputing Centre (http://www.lrz.de) for providing computational resources.

\section*{Supplementary material}
The movies related to this article have been attached as the supplementary material.

\section*{References}
\bibliographystyle{model3-num-names}\biboptions{sort&compress}
\bibliography{aipsamp}
\end{document}